\documentclass [11pt] {article}
\usepackage[dvips]{graphicx,color}
\usepackage[hyperindex,pdfmark,letterpaper]{hyperref}
\newcommand{\bfc}{{\bf c}}
\newcommand{\bfd}{{\bf d}}
\newcommand{\bfF}{{\bf F}}
\newcommand{\bfh}{{\bf h}}
\newcommand{\bfI}{{\bf I}}
\newcommand{\bfu}{{\bf u}}
\newcommand{\bfx}{{\bf x}}

\newcommand{\bftheta}{{\mbox{\boldmath $\theta$}}}
\newcommand{\bfphi}{{\mbox{\boldmath $\phi$}}}
\newcommand{\bfnabla}{{\mbox{\boldmath $\nabla$}}}

\newcommand{\chat}{{\hat{\bfc}}}
\newcommand{\hhat}{{\hat{\bfh}}}
\newcommand{\phihat}{{\hat{\bfphi}}}
\newcommand{\nbar}{{\bar{n}}}
\newcommand{\sigbar}{{\bar{\sigma}}}
\newcommand{\ntil}{{\tilde{n}}}
\newcommand{\util}{{\tilde{u}}}
\newcommand{\bfdtil}{{\tilde{\bfd}}}
\newcommand{\bfutil}{{\tilde{\bfu}}}

\newcommand{\bfxt}{(\bfx, t)}
\newcommand{\bfxp}{(\bfx + \chat_i)}
\newcommand{\bfxtp}{(\bfx, t + 1)}
\newcommand{\bfxpt}{(\bfx + \chat_i, t)}
\newcommand{\bfxmt}{(\bfx - \chat_i, t)}
\newcommand{\bfxptp}{(\bfx + \chat_i, t + 1)}
\newcommand{\eqb}{{\mbox{\scriptsize eq}}}
\newcommand{\nsigi}{{n^\sigma_i}}
\newcommand{\nsigie}{\nsigi^\eqb}
\newcommand{\tausig}{\tau_\sigma}
\newcommand{\taud}{\tau_d}

\usepackage{verbatim}

\begin{document}
\title{
  \begin{flushleft}
    {\footnotesize BU-CCS-990901}\\[0.5cm]
  \end{flushleft}
  \bf A Ternary Lattice Boltzmann Model\\
  for Amphiphilic Fluids}
\author{
  Hudong Chen\\
  {\footnotesize Exa Corporation,}\\
  {\footnotesize Lexington, Massachusetts, U.S.A.}\\
  {\footnotesize {\tt hudong@exa.com}}\\[0.3 cm]
  Bruce M. Boghosian\\
  {\footnotesize Center for Computational Science,}\\
  {\footnotesize Boston University, 3 Cummington Street,
    Boston, Massachusetts 02215, U.S.A.}\\
  {\footnotesize{\tt bruceb@bu.edu}}\\[0.3cm]
  Peter V. Coveney\\
  {\footnotesize Centre for Computational Science,}\\
  {\footnotesize Queen Mary and Westfield College,
    University of London,}\\
  {\footnotesize Mile End Road, London E1 4NS, U.K.}\\
  {\footnotesize{\tt p.v.coveney@qmw.ac.uk}}\\
  }

\date{\today}
\maketitle

\begin{abstract}
  A lattice Boltzmann model for amphiphilic fluid dynamics is presented.
  It is a ternary model, in that it conserves mass separately for each
  chemical species present (water, oil, amphiphile), and it maintains an
  orientational degree of freedom for the amphiphilic species.
  Moreover, it models fluid interactions at the microscopic level by
  introducing self-consistent forces between the particles, rather than
  by positing a Landau free energy functional.  This combination of
  characteristics fills an important need in the hierarchy of models
  currently available for amphiphilic fluid dynamics, enabling efficient
  computer simulation and furnishing new theoretical insight.  Several
  computational results obtained from this model are presented and
  compared to existing lattice-gas model results.  In particular, it is
  noted that lamellar structures, which are precluded by the Peierls
  instability in two-dimensional systems with kinetic fluctuations, are
  not observed in lattice-gas models, but are easily found in the
  corresponding lattice Boltzmann models.  This points out a striking
  difference in the phenomenology accessible to each type of model.
\end{abstract}

\vspace{0.2truein}

\vspace{0.3in}

\pagebreak

\section{Introduction}

Amphiphilic fluids consist of two immiscible phases, together with an
amphiphile species that helps them to mix.  A typical example is a
mixture of oil and water with the addition of detergent.  The molecules
of detergent often have ionic ``heads'' that are hydrophylic, and
hydrocarbon ``tails'' that are hydrophobic, and so there is a strong
energetic preference for these to reside on the oil/water interface.
Moreover, the presence of these orientable particles on the interface
gives it a bending or curvature energy.  These characteristics give
amphiphilic fluids an extremely rich and complicated phenomenology, even
in equilibrium.  A wide variety of structures--including spherical and
wormlike micelles, emulsion droplets, sponge phases and lamellae--can
self-assemble on space and time scales that are very large in comparison
with those of the component molecules, effectively ruling out molecular
dynamics as a suitable methodology for studying such
behaviour~\cite{bib:maillet}.  The addition of fluid dynamics and
nontrivial rheology to such complicated equilibrium behaviour thus
creates a very difficult problem indeed, but one with relevance to a
wide variety of industrial, chemical and biological applications.  These
include oil recovery, pollution remediation, and the dynamics of
vesicles and biomembranes, to name but a few.

Hydrodynamic lattice-gas automata (LGA) were first applied to {\it
  reduced} descriptions of amphiphilic fluids -- in which the amphiphile
species is modelled only by its effects on the surface properties of two
immiscible phases~\cite{bib:turab}.  To capture much of the
above-described phenomenology, however, requires a {\it ternary vector
  model}, in which a separate amphiphile species possesses both
translational and orientational degrees of freedom.  Such models have
been used to study the equilibrium properties of amphiphilic fluids for
more than a decade now~\cite{bib:gomsh}.  A few years ago, it was
demonstrated that hydrodynamic LGA also provide an effective means of
coupling the Hamiltonian of such vector models to hydrodynamic flow with
conserved momentum, thereby providing a self-consistent treatment of the
hydrodynamics of ternary amphiphilic fluids~\cite{bib:bce}.  This model,
which can also be regarded as an extension of Rothman and Keller's
immiscible-fluid algorithm~\cite{bib:roth} to include amphiphilic
species, has been used to study the dynamics of phase
separation~\cite{bib:psep}, the shear-induced sponge-to-lamellar phase
transition~\cite{bib:shear}, and many other phenomena of amphiphilic
fluid dynamics.

In the current work, we create a model of amphiphilic fluid dynamics
that is based on a lattice Boltzmann (LB) model, rather than a LGA.
Both models are kinetic in nature, and both have been implemented in
both two and three dimensions~\cite{bib:metd}, but the LB model is based
on a single-particle distribution function, whereas the LGA is truly
particulate.  Phenomena that depend critically on kinetic fluctuations,
nucleation and other particle discreteness effects are thus well
described by the LGA model, but not by the LB model.  On the other hand,
the LB model is perfectly adequate for the treatment of bulk
hydrodynamic phenomena, is simpler from an algorithmic point of view,
and is more computationally efficient.  Unlike the LGA model, which is
far more complicated in three dimensions than in two owing to its use of
lattice-specific many-body scattering processes, the LB model is easily
implemented on lattices of any dimension.

In addition to introducing the new LB model, an important purpose of the
present work is to understand its limitations relative to the LGA model.
Along these lines, we shall focus on the self-assembly of lamellar
phases in two and three dimensions.  According to a very general theorem
of Peierls~\cite{bib:peierls,bib:ll}, in any infinite system with
underlying kinetic fluctuations, thermodynamic equilibria that are
periodic along one spatial dimension are necessarily unstable in two
dimensions, but may be stable in three dimensions.  Since the LB model
lacks natural kinetic fluctuations, it may be expected to exhibit
lamellae in two dimensions, and we present computer simulation evidence
of this here.  This situation stands in stark contrast to that for the
corresponding LGA model, for which two dimensional lamellae have never
been observed, presumably due to the Peierls instability.

\section{Lattice Boltzmann Models of Hydrodynamics}

LB models come in two varieties: So-called ``top-down''
models~\cite{bib:yso} assume an appropriate form for the free-energy,
and assume that its approach to equilibrium is governed by a
Ginzburg-Landau or Cahn-Hilliard equation; a LB collision operator is
constructed that gives rise to the desired evolution equations in the
hydrodynamic limit.  ``Bottom-up'' models~\cite{bib:shan}, by contrast,
add particle interactions at the kinetic level, and so the form (indeed,
even the existence) of the free energy, and the time-dependence of the
approach to equilibrium are emergent properties of the large scale
dynamics.  Previous LB models of amphiphilic fluid dynamics have all
been ``top-down'' in spirit, with a reduced description using
one~\cite{bib:gomppers} or two~\cite{bib:yeomans} order parameters, but
no surfactant orientational degrees of freedom.  The present model, by
contrast is ternary (it contains three species, although one can also
consider any single or pairwise combination of components) and adopts a
``bottom-up'' approach, making it easier to compare to existing LGA
vector models.  More importantly, unlike either the LGA model or
existing LB models, it offers complete control of the viscosities and
molecular weights of the various species present.  Indeed, the algorithm
allows for any number of different species, including multiple types of
amphiphiles, thus in principle enabling one to study {\it inter alia}
the role of co-surfactants in these systems.  This generality is
evidently beneficial for studying a wider range of physico-chemical
phenomena in amphiphilic fluids.  Moreover, it has previously been
argued that the current approach offers advantages of accuracy in the
resulting hydrodynamic description~\cite{bib:xiaoyi}.

A standard LB system involving multiple species is usually represented
by a set of finite difference equations~\cite{bib:succi,bib:shiyi},
\begin{equation}
  \nsigi\bfxptp - \nsigi\bfxt
  = \Omega^\sigma_i
  \; \; \; \; \; \;
  i = 0, 1, \ldots , b,
  \label{eq:mboz}
\end{equation}
where $\nsigi\bfxt$ is the single-particle distribution function,
indicating the amount of species $\sigma$ (indicating, {\it e.g.}, oil,
water or amphiphile), having velocity $\chat_i$, at site $\bfx$ on a
$D$-dimensional lattice of coordination number $b$, at time step $t$.
The collision operator $\Omega^\sigma_i$ represents the change in the
single-particle distribution function due to collisions.  In the absence
of chemical reactions, this operator must conserve the mass of {\it
  each} species,
\begin{equation}
  \sum_i m_\sigma\Omega^\sigma_i = 0,
\end{equation}
where $m_\sigma$ is the molecular weight for species $\sigma$.
Likewise, in the absence of external forces, it must also conserve the
momentum of {\it all} the species,
\begin{equation}
  \sum_\sigma\sum_i m_\sigma\chat_i\Omega^\sigma_i = 0.
\end{equation}
In the present work, we suppose that the system's microscopic degrees of
freedom are attached to a heat bath at fixed temperature $T$, so that
there is no corresponding requirement for energy conservation.

For convenience, a ``BGK'' form~\cite{bib:krook} is often chosen for the
collision operator,
\begin{equation}
  \Omega^\sigma_i =
  -\frac{1}{\tausig}
  \left(\nsigi - \nsigie\right),
  \label{eq:bgk}
\end{equation}
where the equilibrium distribution function is constructed to satisfy
certain desiderata, including the required conservation
conditions~\cite{bib:chen,bib:qian}.  Most LB models adopt a polynomial
form for this equilibrium distribution function, such as
\begin{equation}
  \nsigie = g_in^\sigma
  \left[
    1 + \beta_0\chat_i\cdot \bfutil +
    \frac{1}{2}\beta^2_0 (\chat_i\cdot \bfutil)^2 -
    \beta_0\frac{\util^2}{2} + \frac{1}{6}\beta^3_0
    (\chat_i\cdot \bfutil)^3 -
    \beta^2_0\frac{\util^2}{2}
    \left(
      \chat_i\cdot \bfutil
    \right)
  \right],
  \label{eq:feq}
\end{equation}
where $n^\sigma\equiv\sum_i n^\sigma_i$, and where the degeneracy
constants $g_i$ and the constant $\beta_0$ are specified entirely by the
choice of lattice.  For example, $\beta_0 = 3$ for the so-called D3Q19
lattice~\cite{bib:qian,bib:fhp2} in three dimensions with 19 states.  In
Eq.~(\ref{eq:feq}) we have defined
\begin{equation}
  \bfutil =
  \frac
  {\sum_\sigma\frac{\rho^\sigma\bfu^\sigma}{\tausig}}
  {\sum_\sigma\frac{\rho^\sigma}{\tausig}}
\end{equation}
where $\rho^\sigma = m_\sigma n^\sigma$ and $\rho^\sigma\bfu^\sigma =
m_\sigma\sum_i\chat_i\nsigi$.  Such a form for the equlibrium
distribution ensures that the local mass and momentum values are
collisionally invariant~\cite{bib:sh_d}.  Indeed, one can directly
verify that
\begin{eqnarray}
  \sum_i\nsigi &=& \sum_i \ntil^\sigma_i,
  \; \; \; \; \; \;
  \forall\sigma\nonumber \\
  \noalign{\noindent and}\\
  \sum_\sigma\sum_i m_\sigma\chat_i\nsigi
  &=& \sum_\sigma\sum_i m_\sigma\chat_i \ntil^\sigma_i,
\end{eqnarray}
where 
\begin{equation}
  \ntil^\sigma_i
  \equiv
  \nsigi + \Omega^\sigma_i
  =
  \nsigi -
  \frac{1}{\tausig}
  \left(\nsigi - \nsigie\right)
\end{equation} 
is the post-collision distribution.  The choice of $\nsigie$ then
results in Navier-Stokes hydrodynamics, where each component $\sigma$
has viscosity $\nu^\sigma = (\tausig - 1/2)/\beta_0$.

\section{Immiscible Lattice Boltzmann Model}

Immiscibility of two or more species is modelled by introducing a
discrete realization of the self-consistently generated mean-field body
force between the particles~\cite{bib:shan}.  Without an amphiphilic
species, and considering only nearest-neighbour interactions among the
immiscible species $\sigma\in\{1,\ldots,\sigma_m\}$, the simplest
possible form for such a force is
\begin{equation}
  \bfF^{\sigma ,c}\bfxt = - \psi^\sigma\bfxt
  \sum_\sigbar\sum_i g_{\sigma\sigbar}
  \psi^\sigbar\bfxpt\chat_i,
  \label{eq:force}
\end{equation}
where the parameter $g_{\sigma\sigbar}$ (positive for repulsive forces,
negative for attractive ones) represents the interaction strength
between the two components, $\sigma$ and $\sigbar$.  Here we have
introduced the {\it effective charge} $\psi^\sigma$, which can be chosen
to be any suitable function of the local mass density $\rho^\sigma$, as
appropriate for the force laws of the hydrodynamic system of interest.
The dynamical effect of the force is realised in the BGK collision
operator in Eq.~(\ref{eq:bgk}) by adding to the velocity $\bfutil$, in
the equilibrium distribution $\nsigie$ of Eq.~(\ref{eq:feq}), an
increment
\begin{equation}
  \delta \bfu^\sigma = \tausig\bfF^{\sigma , c}/\rho^\sigma.
  \label{eq:shift}
\end{equation}
This procedure has recently been shown to be systematically derivable
from the body-force term of the standard continuum Boltzmann equation,
based on a Hermite moment expansion~\cite{bib:martys}.  Indeed, with
this algorithm, phase separation has been demonstrated for sufficiently
large values of $g_{\sigma\sigbar}$~\cite{bib:shan,bib:mart_chen}.  It
is worthwhile to point out that, although the overall momentum change at
any given lattice site and time may be non-zero,
\begin{equation}
  \bfF\bfxt = \sum_\sigma \bfF^{\sigma , c}\bfxt
  = \sum_\sigma \rho^\sigma\bfxt\delta\bfu^\sigma\bfxt,
\end{equation}
Newton's third law, and hence global momentum conservation, is satisfied
exactly by Eq.~(\ref{eq:force}),
\begin{equation}
  \sum_\bfx\bfF\bfxt = 0.
\end{equation}

\section{Amphiphilic Lattice Boltzmann Model}

With the addition of amphiphile, the LB Eqs.~(\ref{eq:mboz}),
(\ref{eq:bgk}), (\ref{eq:feq}) and (\ref{eq:shift}) remain unchanged,
and the species label $\sigma$ simply extends to include the amphiphile
species $s$.  Although as noted above the basic algorithm is general
enough to treat multiple amphiphile species, for simplicity here we
consider only a single type of amphiphile.  Incorporation of the physics
of amphiphilic fluid particles requires two fundamentally new types of
body forces, in addition to those described above.  Unlike ordinary
species such as oil and water, which can be approximated by point (or
spherical) particles so that their interactions depend on relative
distances alone, the essential character of an amphiphilic molecule is
more appropriately represented by a dipole; their interactions depend
not only on their relative distances but also on their dipolar
orientations~\cite{bib:bce}.  It is these additional fundamental
features which result in all the fascinating phenomenology described in
the Introduction.

We first introduce an average dipole vector $\bfd \bfxt$ at each site
$\bfx$ to represent the orientation of any amphiphile present there.
The direction of this dipole vector is allowed to vary continuously in
the present model (although we could do so, we do not attempt to
discretise the dipolar orientation here).  We assume that the dipole
vector for a single amphiphile molecule has magnitude $\hat{d}_0$, and
we suppose that there are $N$ amphiphile molecules represented by a
single lattice-gas particle, so that $|\bfd\bfxt|\leq d_0\equiv
N\hat{d}_0$, where equality holds if and only if all the amphiphile
molecules are perfectly aligned.  Note that the model does not specify
dipole information for each velocity $\chat_i$, for reasons of
computational efficiency and simplicity; information on the average
dipole direction at each site is adequate for the present model.  The
propagation of $\bfd\bfxt$ can then be defined by the simple relation,
\begin{equation}
  n^s\bfxtp\bfd\bfxtp =
  \sum_i\ntil^s_i\bfxmt\bfdtil\bfxmt .
  \label{eq:dprop}
\end{equation}
Note that the {\it dynamics} of the flow is entirely contained in the
distribution function $\ntil^s_i\bfxmt$.  In Eq.~(\ref{eq:dprop}),
$\bfdtil$ represents the post-collision average dipole vector per site,
whose relaxation is also governed by a BGK process,
\begin{equation}
  \bfdtil\bfxt =
  \bfd\bfxt
  - \frac{1}{\taud}
  \left[
    \bfd\bfxt - \bfd^\eqb\bfxt
  \right]
  \label{eq:dbgk}
\end{equation}
with $\taud$ representing relaxation of a dipole vector to a local
equilibrium $\bfd^\eqb\bfxt$.  Eqs.~(\ref{eq:dprop})
and (\ref{eq:dbgk}) completely determine the evolution of the dipoles
once an explicit expression for $\bfd^\eqb\bfxt$ is specified.
Furthermore, it can be seen that the magnitude of the dipole vector is
guaranteed to be less than $d_0$ at all times if $\taud\geq 1$ and
$|\bfd^\eqb|\leq d_0$.

Unlike momentum vectors, there is no conservation requirement associated
for the dipole vectors.  On the other hand, their direction and
magnitude are affected by the distribution of the other dipoles as well
as of oil and water molecules in their neighbourhood.  It is physically
reasonable to assume that the local equilibrium at a given site is the
same for all amphiphile molecules, regardless of their velocity values.
That is, if we let $\bfd^\eqb_i\bfxt$ be the equilibrium dipole value
for amphiphiles of velocity $\chat_i$, then
\begin{equation}
  \bfd^\eqb_i\bfxt = \bfd^\eqb\bfxt,
  \; \; \; \; \; \;
  \forall i.
\end{equation}
Moreover, such an equilibrium property should not depend on the history
of the dipole distribution. Both the orientation and magnitude of the
equilibrium dipole vector are dictated by the nature of the particles in
its neighbourhood and so one of most appropriate choices is to use the
Gibbs measure:
\begin{equation}
  \bfd^\eqb\bfxt 
  = d_0 \frac{\sum_\alpha e^{-\beta H_\alpha\bfxt} 
    \phihat_\alpha} 
  {\sum_\alpha e^{-\beta H_\alpha\bfxt}} 
  \label{eq:ham}
\end{equation}
where $\phihat_\alpha$, for $\alpha\in\{1, \ldots , W\}$ are a set of
unit vectors representing all possible dipole orientations.  Obviously,
Eq.~(\ref{eq:ham}) ensures $|\bfd^\eqb| \leq d_0$.  The parameter
$\beta=1/T$ represents the inverse of the temperature of the
aforementioned heat bath, with which the microscopic relaxation process
is assumed to be in contact.  From Eq.~(\ref{eq:ham}), one can
immediately make the interpretation that the dipole tends to align
itself in the direction which minimizes the energy, $H_\alpha$.  If we
consider the dipoles' vector interactions, it is plausible to assume the
following form for $H_\alpha$, which can be interpreted as a
generalization of the Heisenberg model Hamiltonian in magnetism,
\begin{equation}
  H_\alpha\bfxt
  = -\phihat_\alpha\cdot\bfh\bfxt
\end{equation}
where
\begin{equation}
\bfh\bfxt = \bfh^s\bfxt + \bfh^c\bfxt
\end{equation}
is a mean field created by the surrounding distributions.  Its two
contributions, $\bfh^s\bfxt$ and $\bfh^c\bfxt$, are associated with
distributions of ordinary species and amphiphiles, respectively.  If we
consider only nearest-neighbour interactions for simplicity, these vector
fields have the following forms
\begin{equation}
  \bfh^c\bfxt =
  \sum_\sigma e_\sigma \sum_i n^\sigma\bfxpt\chat_i
  \label{eq:hc}
\end{equation}
where $e_\sigma$ (which may, for example, take its values from the set
$\{-e, 0, +e\}$ where $e$ is constant) is the {\it colour charge} (order
parameter) for ordinary species.  This form may be interpreted
physically as a discrete approximation of the colour gradient of the
immiscible species.  Similarly, we have
\begin{equation}
  \bfh^s\bfxt 
  = \sum_i
  \left[
    \sum_{j \neq 0} 
    n^s_i(\bfx + \chat_j, t)
    \bftheta_j\cdot\bfd_i(\bfx + \chat_j, t) +
    n^s_i\bfxt\bfd_i\bfxt
  \right],
  \label{eq:hs}
\end{equation}
where we have defined the traceless second-rank tensor
\begin{equation}
  \bftheta_j\equiv\bfI - D \frac{\chat_j \chat_j}{c^2},
\end{equation}
and where in turn $\bfI$ is the second-rank unit tensor and $D$ the
spatial dimension.  The physical reason for the appearance of the tensor
$\bftheta_j$ is that dipoles tend to align when their relative positions
are orthogonal to their mean dipolar orientation ({\it i.e.}, $\bfd
\cdot \chat_j = 0$), and anti-align when these are parallel.  Since we
let $\phihat_\alpha$ take on continuous values, in three dimensions
Eq.~(\ref{eq:ham}) can be integrated analytically to give
\begin{equation}
  \bfd^\eqb = d_0
  \left[
    \coth
    \left(
      \beta h
    \right) -
    \frac{1}{\beta h}
  \right]
  \hhat
  \label{eq:deq}
\end{equation}
which becomes
\begin{equation}
  \bfd^\eqb\approx d_0 \frac{\beta h}{3} \hhat
\end{equation}
in the limit of small $\beta h$.  Here we have defined the unit vector
$\hhat\equiv\bfh\bfxt/|\bfh\bfxt|$, and magnitude $h\equiv |\bfh\bfxt|$.

With the presence of an amphiphilic species $s$, the force on an oil or
water particle $\sigma$ needs to include an additional term
\begin{equation} 
  \bfF^\sigma\bfxt = 
  \bfF^{\sigma , c}\bfxt + 
  \bfF^{\sigma , s}\bfxt,
  \label{eq:forc}
\end{equation}
where $\bfF^{\sigma ,c}$ is just the original contribution from the
ordinary molecules given by Eq.~(\ref{eq:force}); the second term in
Eq.~(\ref{eq:forc}) is the contribution from nearby amphiphiles whose
specific form can be derived directly from a leading-order expansion of
Eq.~(\ref{eq:force}). To see how this latter expression may be derived,
recall that an amphiphilic molecule possesses a hydrophilic (positive
colour) head and a hydrophobic (negative colour) tail.  Such a {\it colour
  dipole} can be modelled by a pair of oppositely (colour) charged
fictitious particles with locations displaced by $\pm\bfd/2$ from the
molecular center of mass location $\bfx$.  On this basis, we can derive
an analytical form for the resulting force from Eq.~(\ref{eq:force}) via
a Taylor expansion in $\bfd$.  When only nearest-neighbour interactions
are considered, the form of this interaction is given by
\begin{equation}
  \bfF^{\sigma , s}\bfxt
  = - 2\psi^\sigma\bfxt g_{\sigma s}
  \sum_{i \neq 0} \bfdtil\bfxpt
  \cdot\bftheta_i
  \psi^s\bfxpt,
  \label{eq:bscn}
\end{equation}
where the parameter $g_{\sigma s}$ is the coupling coefficient between
an ordinary and an amphiphilic species.

Using similar physical arguments, we can derive the forces on
amphiphilic particles,
\begin{equation} 
  \bfF^s\bfxt = 
  \bfF^{s, c}\bfxt + \bfF^{s, s}\bfxt,
\end{equation}
where $\bfF^{s ,c}$ and $\bfF^{s ,s}$ represent forces from ordinary
species and amphiphiles, respectively.  In similar fashion, explicit
forms for both of these forces can be derived directly from the force
law, Eq.~(\ref{eq:force}), for ordinary immiscible species; physically,
the force $\bfF^{s ,c}$ is simply the reaction force to $\bfF^{\sigma ,
  s}$.  It may be shown to have the following form,
\begin{equation}
  \bfF^{s, c}\bfxt
  = 2\psi^s\bfxt \bfdtil\bfxt
  \cdot \sum_\sigma g_{\sigma s}
  \sum_{i \neq 0}\bftheta_i
  \psi^\sigma\bfxpt,
  \label{eq:bcsn} 
\end{equation}
from which we can verify that Newton's third law is satisfied,
\begin{equation}
  \sum_\bfx
  \left[
    \sum_\sigma \bfF^{\sigma , s}\bfxt +
    \bfF^{s, c}\bfxt
  \right] = 0.
\end{equation}

To the leading order of a Taylor expansion in the ratio of
$\left|\bfc_i\right|$ to the colour gradient scale length,
Eqs.~(\ref{eq:bscn}) and (\ref{eq:bcsn}) reduce to
\begin{equation}
  \bfF^{\sigma , s}(\bfx)
  = -\frac{2bc^2}{D(D+2)} g_{\sigma s} \psi^\sigma (\bfx)
  \left\{
    \nabla^2\left[\bfdtil(\bfx)\psi^s(\bfx)\right] -
    D\bfnabla\bfnabla\cdot\left[\bfdtil(\bfx)\psi^s(\bfx)\right]
  \right\}
\end{equation}
and
\begin{equation}
  \bfF^{s, c}(\bfx)
  = \frac{2bc^2}{D(D+2)} g_{\sigma s}\psi^s (\bfx)\bfdtil(\bfx)
  \cdot\sum_\sigma
  \left[
    \bfI\nabla^2\psi^\sigma (\bfx) -
    D\bfnabla\bfnabla\psi^\sigma (\bfx)
  \right].
\end{equation}
The above expressions make manifest the physically reasonable result
that fluid particles experience no net force if the other particles in
their neighorhood are distributed uniformly.  More interesting is the
result that the leading-order contributions involve second-order spatial
derivatives.  This implies that amphiphiles tend to reside near the
interface between two immiscible fluids.

Finally, $\bfF^{s, s}$ represents the force due to interactions between
amphiphilic particles.  This depends not only on the relative distance
but also on the relative orientation of the dipoles in question.  Though
considerably more involved algebrically, its expression is still
derivable from small expansions of the polarization $\bfd$ about $\bfx$
as well as those at the neighbouring sites, as described above.  In
particular, when only nearest-neighbour interactions are considered, this
force has the simple form
\begin{eqnarray}
  \bfF^{s,s}(\bfx) &=& - \frac{4D}{c^2}
  g_{ss} \psi^s(\bfx) \sum_i
  \left\{
    \bfdtil\bfxp\cdot\bftheta_i\cdot\bfdtil(\bfx)\chat_i
  \right.
  \nonumber \\
  & & +
  \left.
    \left[
      \bfdtil\bfxp\bfdtil(\bfx) +
      \bfdtil(\bfx) \bfdtil\bfxp
    \right]
    \cdot \chat_i
  \right\}
  \psi^s\bfxp.
  \label{eq:bssn}
\end{eqnarray}
Here the coupling coefficient $g_{ss}$ should be negative if we wish to
model attraction between two amphiphile heads and repulsion between a
head and a tail.  Likewise, it can be verified directly that global
momentum conservation is satisfied by this form; that is,
\begin{equation}
  \sum_\bfx \bfF^{s,s}\bfxt = 0,
\end{equation}
when boundary effects are absent.

To summarize, the basic LB evolution equation is given by
Eq.~(\ref{eq:mboz}).  The BGK operator for collisional relaxation is
given by Eq.~(\ref{eq:bgk}), where the collisional equilibrium is given
by Eq.~(\ref{eq:feq}); this describes an ordinary LB algorithm for
miscible Navier-Stokes fluids.  To realise immiscibility, we implement
the force, Eq.~(\ref{eq:force}), between neighbouring particles by
shifting the velocity used in Eq.~(\ref{eq:feq}) by the amount specified
in Eq.~(\ref{eq:shift}); this describes the previously known
``bottom-up'' immiscible fluid LB algorithm. The principal contribution
of this work is the introduction of a surfactant species with an
orientation that propagates according to Eq.~(\ref{eq:dprop}), and
relaxes according to the BGK-like orientational collision operator,
Eq.~(\ref{eq:dbgk}), with orientational equilibrium given by
Eqs.~(\ref{eq:deq}), (\ref{eq:hc}) and (\ref{eq:hs}).  The force on
ordinary oil/water species due to amphiphile is given by
Eq.~(\ref{eq:bscn}), that on amphiphile due to ordinary species is given
by Eq.~(\ref{eq:bcsn}), and that on amphiphile due to other amphiphile
is given by Eq.~(\ref{eq:bssn}).  The effect of these forces on the
collisional relaxation is also treated by shifting the velocity used in
the BGK equilibrium by an amount specified in Eq.~(\ref{eq:shift}).
Taken together, these equations completely specify the new ternary
amphiphilic LB model and associated algorithm.

\section{Simulations}

In this section we demonstrate qualitatively that some of the basic
phenomenology of amphiphilic fluids is indeed exhibited by the new
model.  We begin by verifying that the model with the amphiphilic
interactions extinguished displays complete oil-water phase separation.
All simulations shown in this section were performed on a domain of size
$32\times 32$ lattice units with periodic boundary conditions, and used
$\tau_\sigma=\tau_s=\tau_d=1$ and $\beta=10$.

\begin{figure}
  \begin{center}
    \mbox{
      \includegraphics[bbllx=142,bblly=196,bburx=470,bbury=595,width=0.4\textwidth]{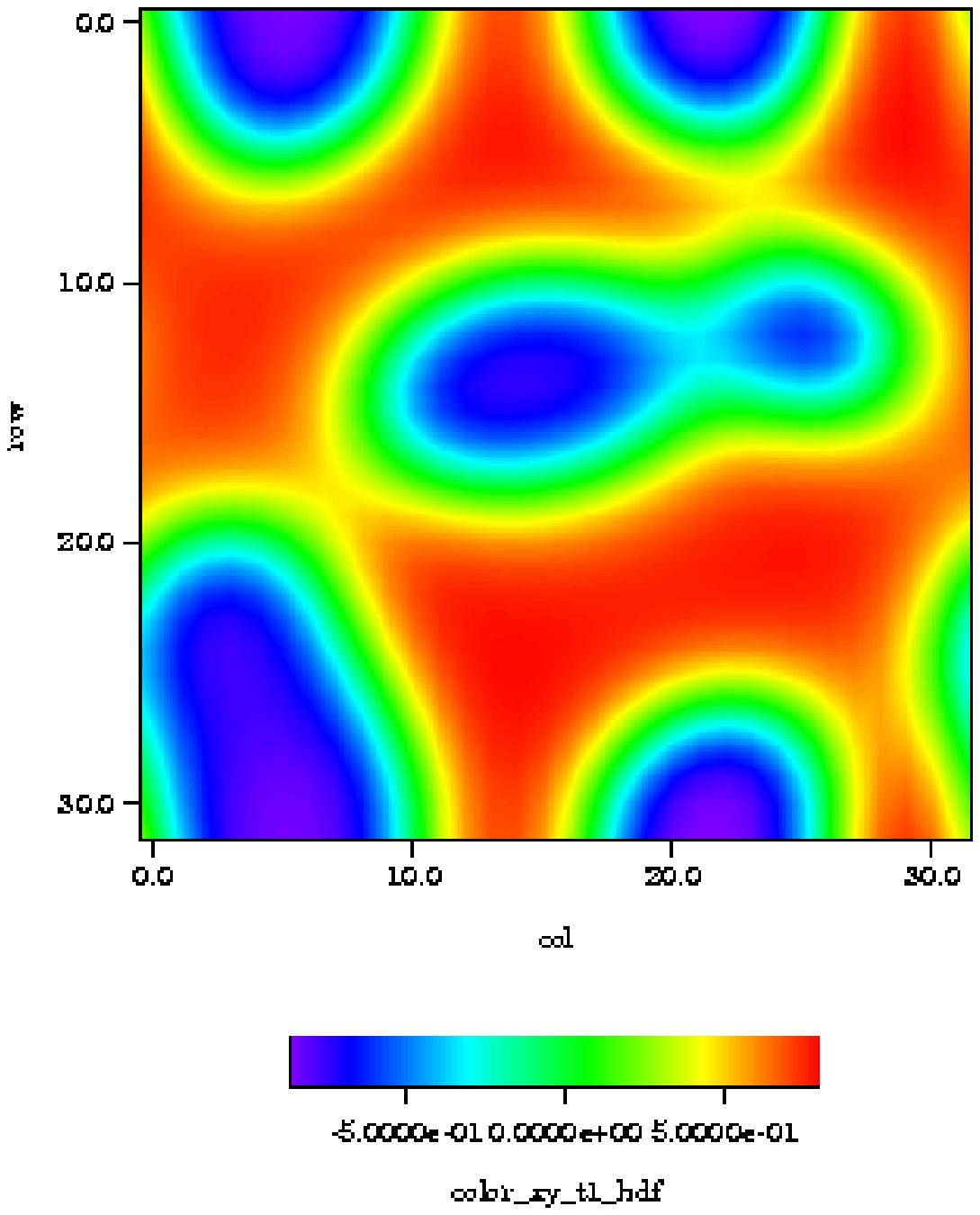}
      \includegraphics[bbllx=142,bblly=140,bburx=470,bbury=595,width=0.4\textwidth]{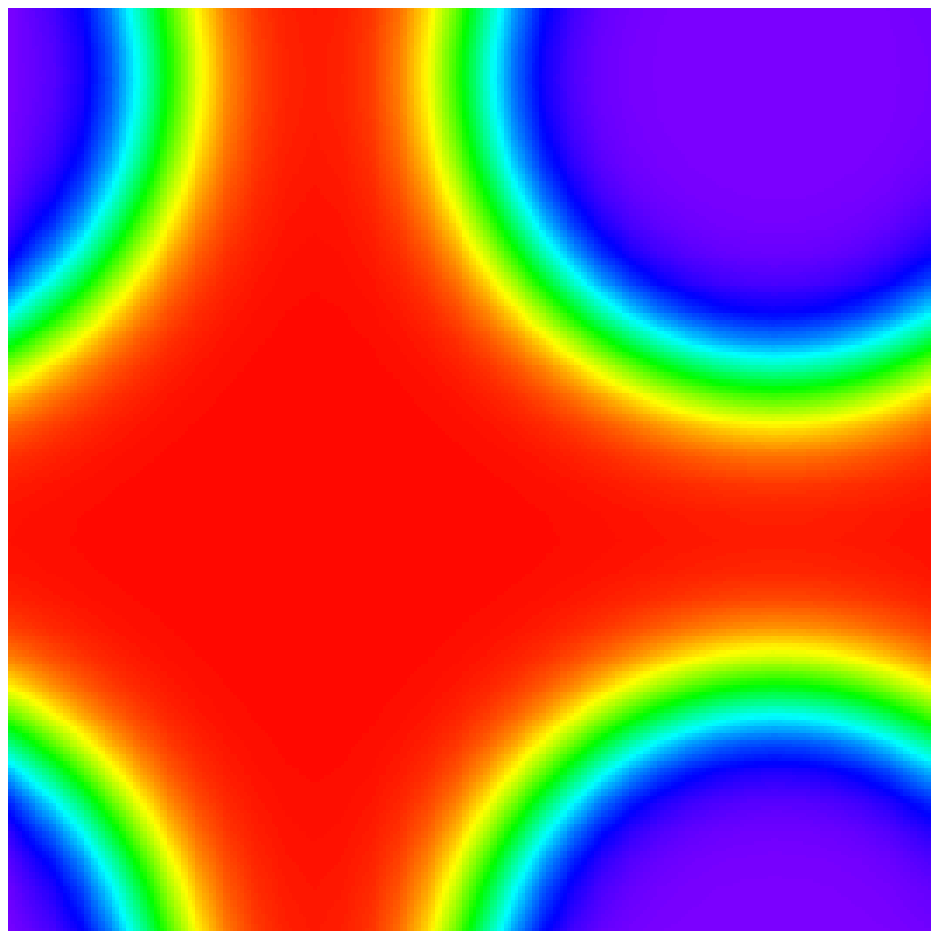}
      }
  \end{center}
  \caption{Two-dimensional binary oil(red)-water(blue) phase separation. 
   Colour (oil-water) distributions shown at lattice timesteps 
   $t = 6000$ (left) and $t = 16000$ (right)}
  \label{fig:sep}
\end{figure}

Fig.~\ref{fig:sep} shows the equilibrated colour order parameter at 6000
and 16000 time steps after an off-critical quench.  The run began with
blue density $\nbar^b = 0.45$, red density $\nbar^r = 0.65$ and
surfactant density $\nbar^s = 0.25$, distributed homogeneously in space,
aside from a small random fluctuation.  For sufficiently strong
interaction parameter ($g_{br} = 0.03$ was used here) between the
immiscible red and blue phases, complete separation was observed as can
be seen in the figure.  Because the amphiphile interaction parameters
were turned off ($g_{bs} = g_{ss} = 0$), the separation did not arrest,
and the blue fluid formed a single droplet (recall that the boundary
conditions are periodic).

\begin{figure}
  \begin{center}
    \mbox{
      \includegraphics[bbllx=142,bblly=196,bburx=470,bbury=595,width=0.4\textwidth]{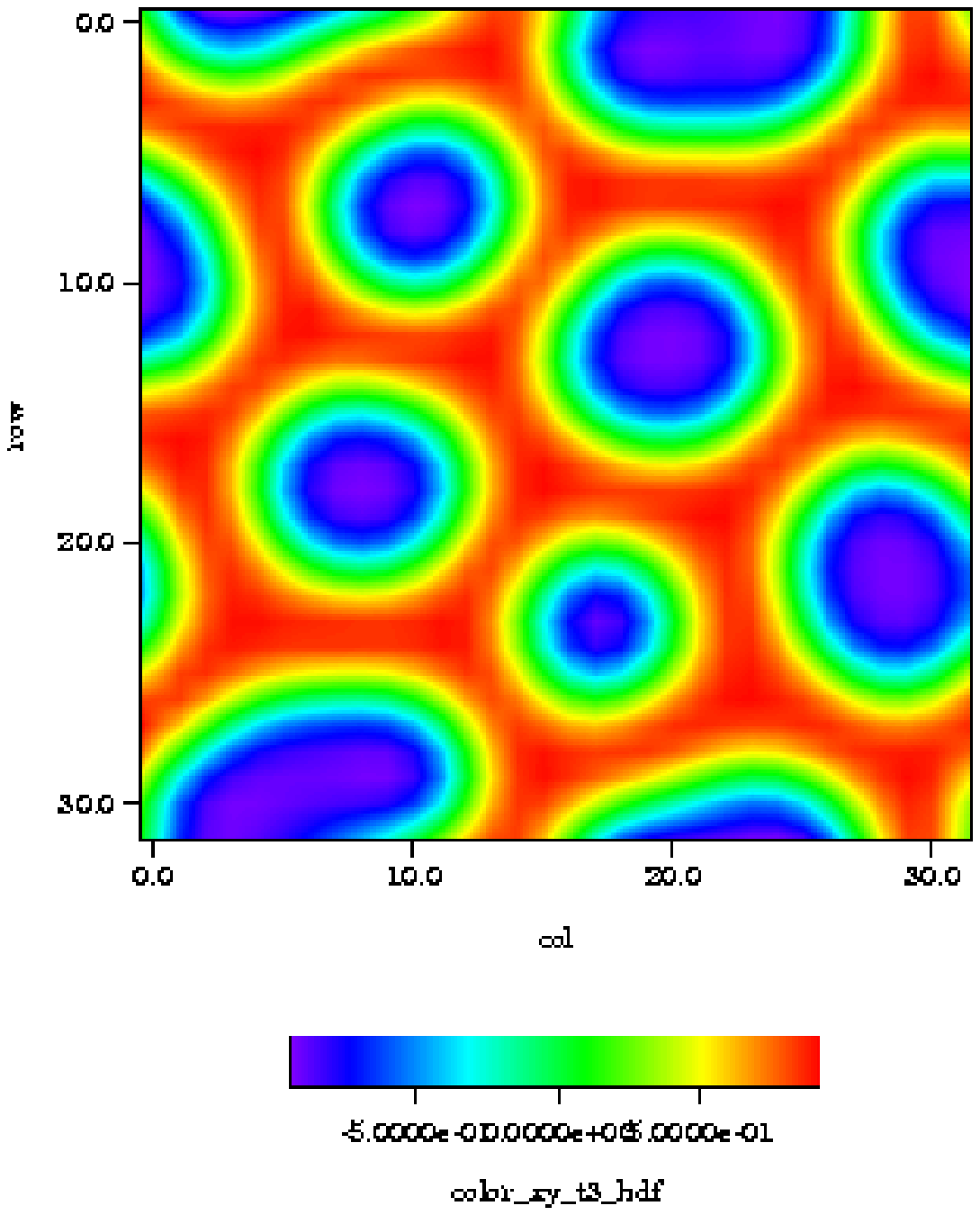}
      \includegraphics[bbllx=142,bblly=170,bburx=470,bbury=595,width=0.4\textwidth]{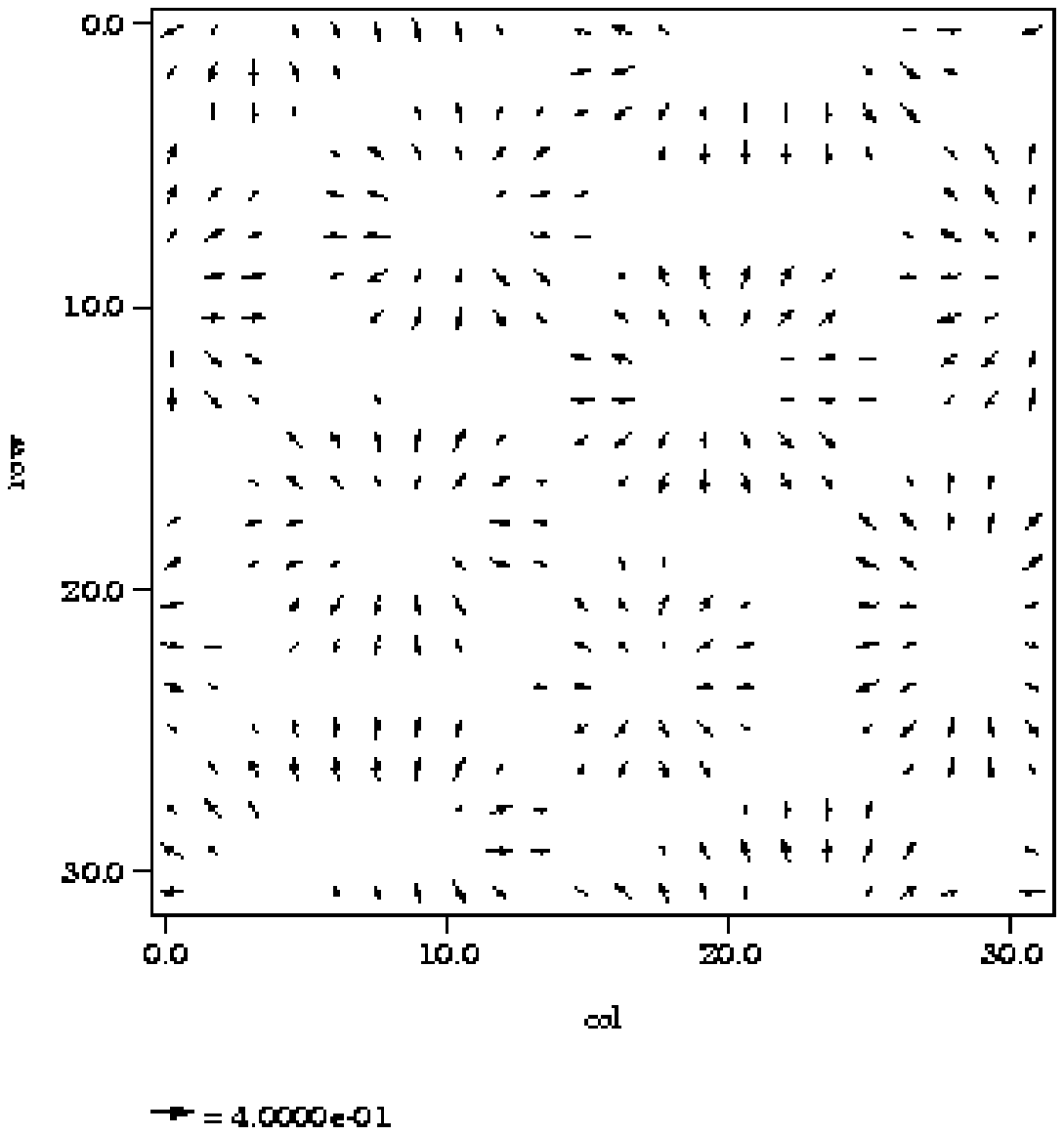}
      }
  \end{center}
  \caption{Two-dimensional domain growth in a ternary
    (oil-water-surfactant) fluid. Colour (oil-water, left) and 
    amphiphile director (right) distributions are both shown at 
    lattice timestep $t = 16000$}
  \label{fig:arrest}
\end{figure}

Next, using the same density values and initial conditions as above, but
with the amphiphilic interactions switched on ($g_{bs} = g_{ss} = -
0.01$), Fig.~\ref{fig:arrest} shows the colour order parameter and the
surfactant director density after 16000 time steps, by which time the
system appears to be in a stationary state.  Note that the amphiphile
has organised itself on the surface, with directors pointing from blue
to red on average.  This effectively demonstrates that the amphiphile
has arrested the phase separation; if the separation were to continue,
there would not be sufficient surface area to accommodate the
amphiphile, which incurs a large energy penalty for being away from a
surface.  Such arresting of domain growth has previously been observed
in LGA studies of amphiphilic fluids~\cite{bib:psep}.

\begin{figure}
  \begin{center}
    \mbox{
      \includegraphics[bbllx=142,bblly=196,bburx=470,bbury=595,width=0.4\textwidth]{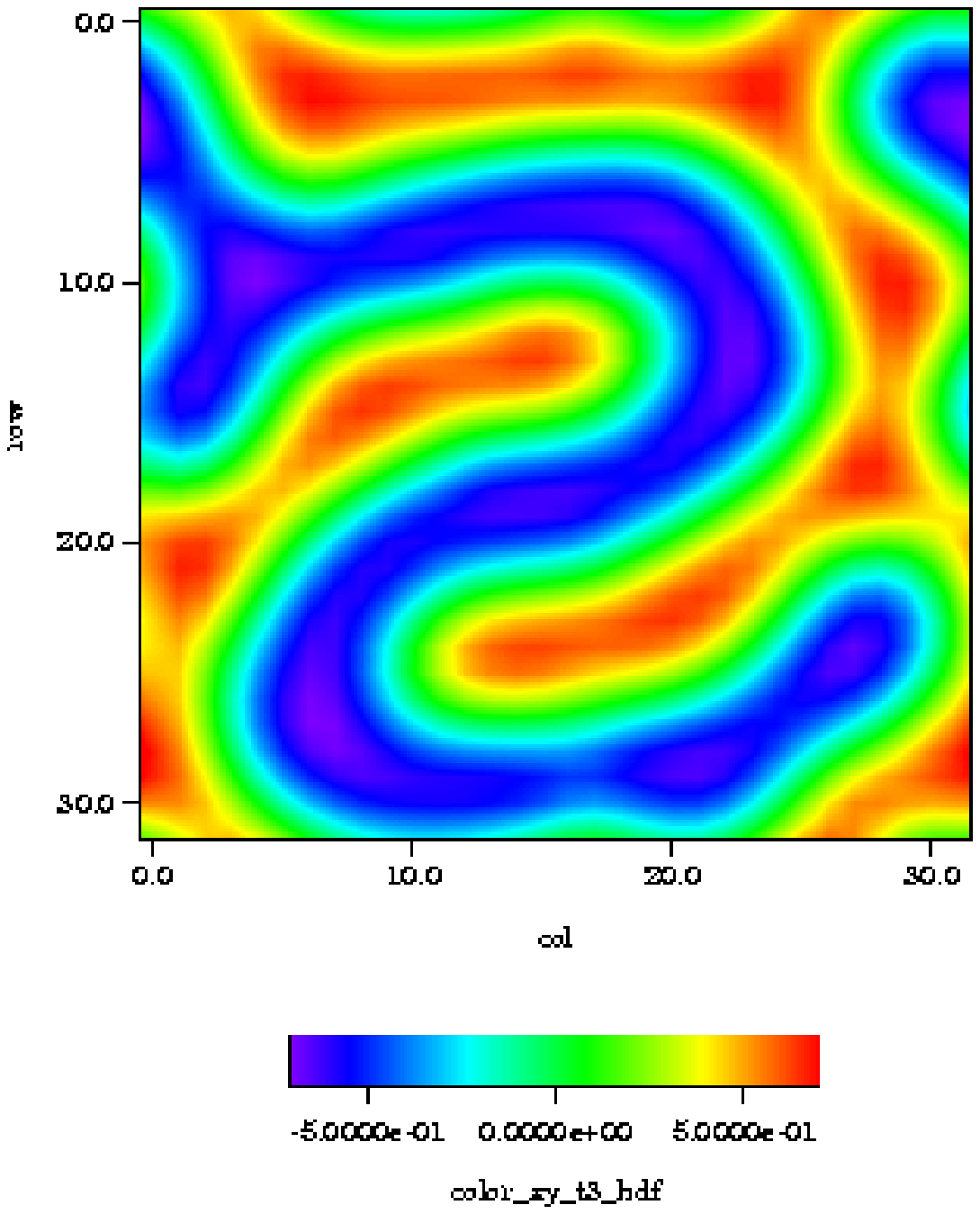}
      \includegraphics[bbllx=142,bblly=170,bburx=470,bbury=595,width=0.4\textwidth]{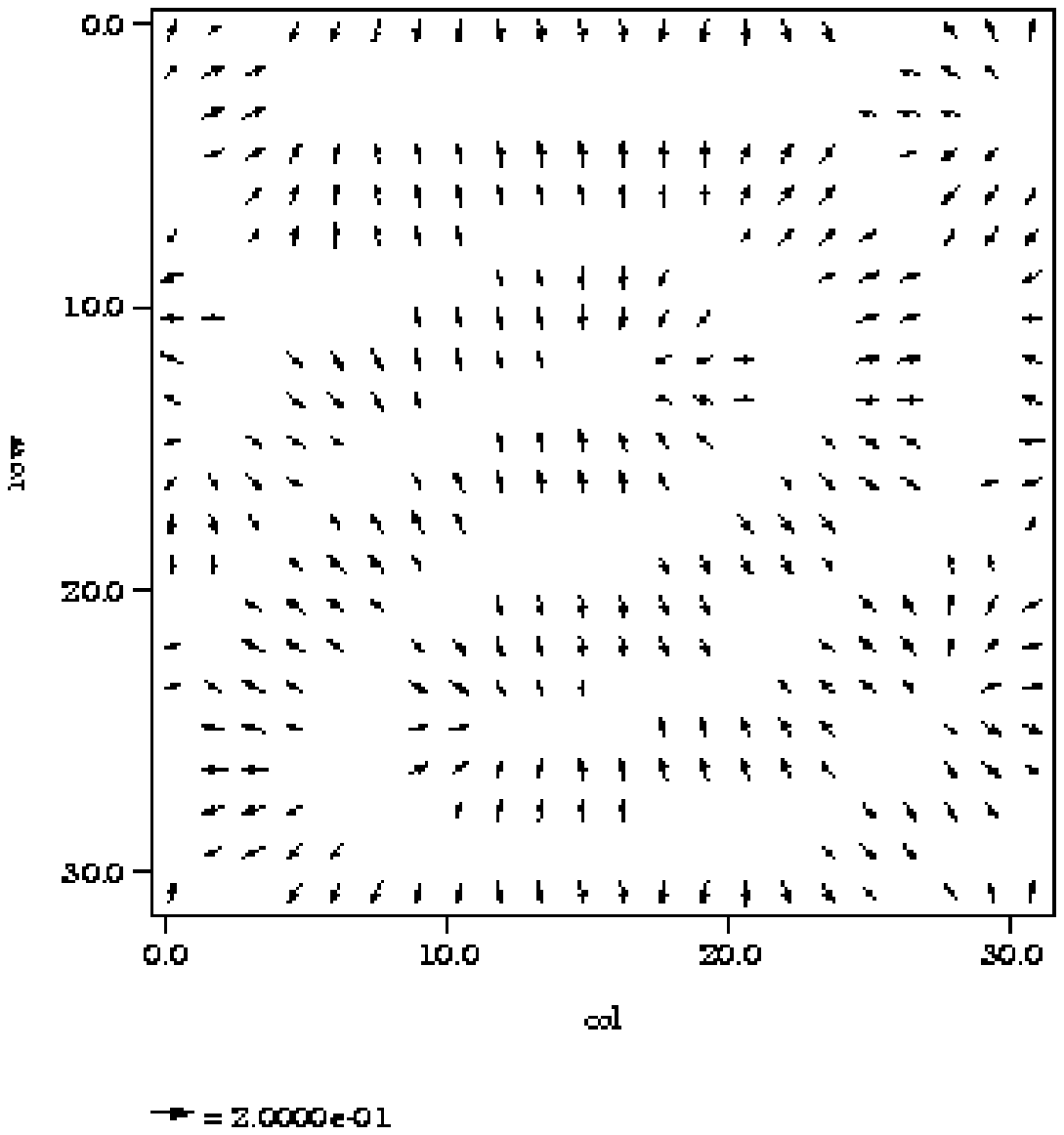}
      }
  \end{center}
  \caption{Formation of a lamellar phase in a ternary oil-water
    surfactant lattice Boltzmann fluid. Colour (left) and amphiphile 
    (right) distribution are shown, both 
    at lattice timestep $t = 20000$}
  \label{fig:lam}
\end{figure}

Fig.~\ref{fig:lam} shows the equilibrated colour order parameter and
director density, 20000 time steps after a critical quench on a domain
of size $32\times 32$ lattice units.  The run began with $\nbar^b =
\nbar^r = 0.5$ and $\nbar^s = 0.1$, distributed homogeneously in space,
aside from a small random fluctuation.  No further evolution of the
order parameter was observed after a few thousand time steps,
strengthening the case that this is indeed an equilibrium state.

\section{Discussion and Conclusions}

As mentioned in the Introduction, an important focus of our studies to
date with the new model has centered on the question of the existence of
lamellae in two spatial dimensions.  The theorem that we invoke was
derived by Peierls in the context of electronic
structure~\cite{bib:peierls,bib:ll}, but is general enough to apply
here.  Let us suppose that a structure $\phi(x)$ that is periodic along
dimension $x$ minimizes some free energy functional $F[\phi]$.  Let us
further suppose that this structure undergoes thermally induced
fluctuations in a Gibbsian distribution $\exp(-\delta F/T)$, where
$\delta F$ is the free energy excess due to the fluctuation.  Peierls
demonstrated that the variance of the component of such fluctuations in
the direction of the periodicity is proportional to
\begin{equation}
T\int\frac{dk_x\;k_\perp dk_\perp}{\phi(k_x, k_\perp^2)},
\end{equation}
where $\phi$ is an analytic function of its arguments that vanishes when
its arguments are zero.  This variance is easily seen to diverge
logarithmically for small wavevector (large system size) for any finite
temperature $T>0$.  Moreover, similar arguments show that such a
divergence does not occur in three spatial dimensions.

Thus, we should expect lamellar structures to be unstable in an infinite
two dimensional system with finite-temperature fluctuations.  Indeed,
previous lattice-gas studies of amphiphilic fluids in two
dimensions~\cite{bib:bce,bib:shear} have not found stable lamellae,
despite extensive parametric searches for them, except when the system
was sheared~\cite{bib:shear}.  Lattice-gas studies in three dimensions,
however, do exhibit stable unsheared lamellar
structures~\cite{bib:metd}.  We suggest here that the observed
instability of lamellae in two dimensions is due to the Peierls
mechanism.  However, we must qualify this claim with the proviso that it
is somewhat dangerous to assume that the Peierls instability applies in
a lattice-gas simulation: although a LGA is indeed a finite-temperature,
fluctuating system, computer simulations of it always have a finite
domain size, and hence a minimum sustainable wavevector. This objection
might be mitigated if it could be shown that lamellar phases reappear at
zero temperature.  Unfortunately, it is impossible to lower the
effective temperature of a LGA to zero; kinetic fluctuations will always
be present by the very nature of the model.  LB methods, by contrast,
have no kinetic fluctuations.  Thus, if an LB model were to exhibit
lamellae in two dimensions, it would strengthen the argument that their
absence in the LGA model is due to the Peierls instability.  As
described in the previous section, we tested this by simulating the
critical quench shown in Fig.~\ref{fig:lam}; that this resulted in a
lamellar phase supports the above interpretation of the Peierls
instability.

In conclusion, we have described a ternary lattice Boltzmann model of
amphiphilic fluids.  This model differs from previous LB models in that
it accounts for amphiphilic director orientation, and incorporates
microscopic interactions between the fluid particles, without the need
to postulate a free energy functional.  It differs from LGA models in
that it does not include fluctuations associated with particle
discreteness.  The lack of fluctuations gives the model a different
phase behaviour from the corresponding LGA; in particular, we have
presented evidence from computer simulation that it exhibits lamellar
phases in two dimensions, which are prohibited by the Peierls
instability in models with fluctuations.

\section*{Acknowledgements}

The authors would like to thank Ye Zhou for his help with coding an
earlier version of the model.  The authors' collaboration was
facilitated by a travel grant from NATO (grant number CRG950356).  BMB
was supported in part by the United States Air Force Office of
Scientific Research under grant number F49620-95-1-0285, and in part by
AFRL.


\begin{thebibliography}{99}
  \medskip
  
\bibitem{bib:maillet} J-B. Maillet, V. Lachet and P.V. Coveney ``Large
  scale molecular dynamics simulation of self-assembly processes in
  short and long chain cationic surfactants'', {\it Phys. Chem. Chem.
    Phys.}, {\bf 1} (1999), in press.
  
\bibitem{bib:turab} S-Y. Chen, T. Lookman, {\it J. Stat. Phys.}, {\bf
    81}, 223-235 (1995).
  
\bibitem{bib:gomsh} G. Gompper, M. Schick, {\it Phase Transitions and
    Critical Phenomena}, {\bf 16} (1994) 1-181.
  
\bibitem{bib:bce} B. Boghosian, P. Coveney, and A. Emerton, {\em Proc.
    R. Soc. Lond. A}, {\bf 452}, 1221 (1996).
  
\bibitem{bib:roth} D. H. Rothman and J. M. Keller, {\em J. Stat. Phys.,}
  {\bf 52}, 1119 (1988); A. K. Gunstensen, D. H. Rothman, S.
  Zaleski, and G. Zanetti, {\em Phys. Rev. A}, {\bf 43}, 4320 (1991).
  
\bibitem{bib:psep} A.N. Emerton, P.V. Coveney, B.M. Boghosian, {\it
    Phys. Rev. E}, {\bf 55}, 708-720 (1997).
  
\bibitem{bib:shear} A.N. Emerton, F.W.J. Weig, P.V. Coveney, 
    B.M. Boghosian, {\it J. Phys. Condens. Mat.}, {\bf 9} 8893-8905 
    (1997).

\bibitem{bib:metd} B.M. Boghosian, P.V. Coveney and P.J. Love,
  ``Three-dimensional lattice-gas model for amphiphilic fluid
  dynamics'', {\it Proc. R. Soc. London A}, in press (1999).

\bibitem{bib:peierls} R.E. Peierls, ``Quantum Theory of Solids,''
  Clarendon Press, Oxford University (1955).  
  
\bibitem{bib:ll} L.D. Landau and E.M. Lifshitz, ``Statistical Physics,
  Part I,'' Butterworth Heinemann (1997), pp. 432-436.

\bibitem{bib:yso} M. Swift, W. Osborn and J. Yeomans, {\em Phys.
    Rev. Lett.}, {\bf 75}, 830 (1995); M. Swift, S. Orlandini, W.
  Osborn and J. Yeomans, {\em Phys. Rev. E}, {\bf 54}, 5041 (1996).
  
\bibitem{bib:shan} X. Shan and H. Chen, {\em Phys. Rev. E}, {\bf 47},
  1815 (1993); X. Shan and H. Chen, {\em Phys. Rev. E}, {\bf 49},
  2941 (1994).
  
\bibitem{bib:gomppers} O. Theissen, G. Gompper and
    D.M. Kroll, {\it
    Europhys. Lett.}, {\bf 42}, 419 (1998).
  
\bibitem{bib:yeomans} G. Gonella, E. Orlandini and J.M. Yeomans, {\it
    Phys. Rev. Lett.}, {\bf 78}, 1695 (1997); A. Lamura, G. Gonella
    and J.M. Yeomans, {\it Int. J. Mod. Phys. C}, {\bf 9}, 1469 (1998).
  
\bibitem{bib:xiaoyi} X. He, X. Shan, and G. Doolen, {\em Phys. Rev. E},
  {\bf 57}, R13 (1998).
  
\bibitem{bib:succi} F. Higuera, S. Succi and R. Benzi, {\em Europhys.
    Lett.}, {\bf 9}, 345 (1989); and R. Benzi, S. Succi and M.
  Vergassola, {\em Physics Reports}, {\bf 222}, 145 (1992).
  
\bibitem{bib:shiyi} S. Chen and G. Doolen, {\em Ann. Rev. Fluid Mech},
  {\bf 30}, 329 (1998).
  
\bibitem{bib:krook} P. Bhatnagar, E. Gross and M. Krook, {\em Phys.
    Rev.}, {\bf 94}, 511 (1954).
  
\bibitem{bib:chen} S. Chen, H. Chen, D. Martinez and W. Matthaeus, {\em
    Phys. Rev. Lett.}, {\bf 67}, 3776 (1991); H. Chen, S. Chen and W.
  Matthaeus, {\em Phys. Rev. A}, {\bf 45}, 5339 (1992).
  
\bibitem{bib:qian} Y. Qian, D. d'Humi{\` e}res and P. Lallemand, {\em
    Europhys. Lett.}, {\bf 17}, 479 (1992).
  
\bibitem{bib:fhp2} U. Frisch, D. D'Humieres, B. Hasslacher, P.
  Lallemand, Y. Pomeau, and J. Rivet, {\em Complex Systems}, {\bf 1},
  649 (1987).
  
\bibitem{bib:sh_d} X. Shan and G. Doolen, {J. Stat. Phys.}, {\bf 81},
  379 (1995).
  
\bibitem{bib:martys} N. Martys, X. Shan, and H. Chen, {\em Phys. Rev.
    E}, {\bf 58}, 6855 (1998).
  
\bibitem{bib:mart_chen} N. Martys and H. Chen, {Phys. Rev. E}, {\bf 53},
  743 (1995).
  

\end{thebibliography}
\end{document}